\documentclass[%
 preprint,
nofootinbib,
 amsmath,amssymb,
 aps,
 prl,
 longbibliography,
 lengthcheck,%
]{revtex4-1}

\usepackage{graphicx}% Include figure files
\usepackage{dcolumn}% Align table columns on decimal point
\usepackage{bm}% bold math
\usepackage{hyperref}% add hypertext capabilities
\usepackage{mathtools}
\usepackage[mathlines]{lineno}% Enable numbering of text and display math
\usepackage{changepage} % for the adjustwidth environment
\usepackage{xcolor}
\usepackage{soul}
\newcommand{\be}{\begin{equation}}
\newcommand{\ee}{\end{equation}}

\newcommand{\la}{\langle}
\newcommand{\ra}{\rangle}

\newcommand{\cH}{{\cal H}}
\newcommand{\cS}{{\cal S}}
\newcommand{\cB}{{\cal B}}

\newcommand{\densS}{ \rho^{\cal S}}

\newcommand{\meandensS}{\overline{{ \rho}^{\cal S}}}

\newcommand{\tr}{\mathrm{tr}}

\newcommand{\ii}{\mathrm{i}}

\begin{document}

%\linenumbers

\title{
Testing eigenstate decoherence hypothesis in a model of collisional decoherence
}

\author{Ivan V. Dudinets$^{1,2}$, Igor Ermakov$^{1,2,3}$,  Oleg Lychkovskiy$^{1,2,3}$}

\affiliation{$^1$Skolkovo Institute of Science and Technology,
Skolkovo Innovation Center 3, Moscow  143026, Russia}
\affiliation{$^2$Moscow Institute of Physics and Technology, Institutsky per. 9, Dolgoprudny, Moscow  region,  141700, Russia}
\affiliation{$^3$Department of Mathematical Methods for Quantum Technologies, Steklov Mathematical Institute of Russian Academy of Sciences,
Gubkina str. 8, Moscow 119991, Russia}

\date{\today}% It is always \today, today,


\begin{abstract}
The eigenstate decoherence hypothesis (EDH) asserts that each individual eigenstate of a large closed system is locally classical-like. We test this hypothesis for a heavy particle interacting with a gas of light particles. This system is paradigmatic for studies of the quantum-to-classical transition: The reduced state of the heavy particle is widely believed to rapidly loose any nonclassical features due to the interaction with the gas. Yet, we find numerical evidence that the EDH is violated: certain eigenstates of this model are manifestly non-classical. Only the weak version of EDH referring to the majority (instead of the totality) of eigenstates holds.
\end{abstract}

\maketitle

\noindent{\it Introduction.} Quantum mechanical state space is dramatically different from the classical state space. For example, a state of a point particle in classical mechanics is completely determined by its coordinate and momentum, and thus the state space is the two-dimensional phase space. In quantum mechanics the coordinate and momentum can not be determined simultaneously.  However, the superposition principle of quantum theory legitimizes arbitrary superpositions of different coordinates (or momenta). As a result, the state space becomes an infinite-dimensional projective Hilbert space. Most of the points in the Hilbert space represent weird superpositions that do not make any sense from the classical point of view. These weird states apparently do not emerge in our everyday experience (and for this reason do not enter the classical formalism). Explaining why this happens is the key for understanding the quantum-to-classical transition.

A conceptually straightforward explanation is provided by the  decoherence theory \cite{zeh1970interpretation,zeh1973toward,zurek1981pointer,zurek1982environment,joos1985emergence,schlosshauer2008decoherence}. One observes that the weird superposition states turn out to be extremely fragile when the interaction between the system and its environment is taken into account. To demonstrate this fact one often considers a system  $\cS$ (e.g. a point particle) and its environment (bath) $\cB$ (e.g. the gas the particle is immersed in) as a combined closed systems $\cH$, whose state $\Psi_t$ evolves according to the Schr\"odinger equation
$
i \partial_t \Psi(t) = H \Psi(t),
$
$H$ being the total Hamiltonian that describes the system, the environment and their interaction (we adopt the convention $\hbar=1$).  The state of the (now open) system $\cS$ is described by the reduced density matrix
\be\label{densS}
\densS(t)\equiv \tr_{\cB} |\Psi(t)\ra\la\Psi(t)|,
\ee
where $\tr_{\cB}$ is the partial trace over the environment $\cB$.
It has been verified in various physical settings that even when $\densS(0)$ is a projection on a highly non-classical state, $\densS(t)$ becomes a mixture of classical-like states on a certain {\it decoherence time scale}, the latter being  extremely  short for a system $\cS$ that can be regarded as macroscopic (e.g. for a macroscopic point particle) \cite{schlosshauer2008decoherence}.

Importantly, the focus of the above-described studies in the framework of the decoherence theory has always been on out-of-equilibrium initial states $\Psi(0)$. In particular, often $\Psi(0)$ of the product form has been considered, describing initially uncorrelated system and environment. A complementary view can be provided by considering the {\it eigenstates} of the combined system $\cH$. They remain unchanged under the Schr\"odinger equation (apart from the irrelevant phase factor),  and one naturally expects that they are locally classical-like from the outset. This is the essence of the {\it eigenstate decoherence hypothesis} (EDH) put forward by one of us some time ago \cite{lychkovskiy2013dependence}. More formally, the EDH asserts that for any eigenstate $\Phi_E$ of the total Hamiltonian $H$ the corresponding reduced density matrix
\be
\densS_E \equiv \tr_{\cB}|\Phi_E\ra\la \Phi_E|
\ee
is classical-like according to a suitable quantumness measure, or, simply put,  free of any weird
quantum Schr\"odinger-cat-type superpositions.

The EDH is motivated by the eigenstate thermalization hypothesis (ETH) \cite{deutsch1991quantum,srednicki1994chaos,rigol2008thermalization}. The later asserts that $\densS_E$ is a smooth function of $E$. This implies, by standard statistical arguments, that $\densS_E$ is thermal, and that a long-time average of any evolving state with subextensive energy uncertainty is thermal  \cite{Mori_2018}. Various case studies have suggested that $\densS_E$ indeed has all attributes of an equilibrium state for generic nonintegrable systems without disorder  \cite{rigol2008thermalization,kim2014testing,beugeling2014finite-size}. On the other hand, it has been recently realized that many nonintegrable systems possess {\it quantum many-body scars}, i.e.  abnormal states in the bulk of the otherwise thermal spectrum that violate the ETH \cite{shiraishi2017systematic,turner2018weak,Moudgalya_2018_exact}.

One reasonably expects that if a state is thermal, the more so it should be classical-like \cite{joos1985emergence}. Combining this straightforward idea with the ETH, one immediately arrives to the EDH \cite{lychkovskiy2013dependence}. Furthermore, one can expect that the EDH should be valid for a larger class of systems than the ETH, e.g. for interacting integrable systems \cite{lychkovskiy2013dependence} and system with quantum many-body scars. This expectation is rooted in the fact that integrability or presence of scars is a global property sensitive e.g. to long range interactions, and a system is not able to explore whether it is integrable, scarred or fully ergodic on an extremely short decoherence time scale.
%Note that the reverse is not necessarily true: most of the classical-like states are out of equilibrium.

The EDH has not been tested in specific models up to now, with the exception of a central spin model \cite{lychkovskiy2013dependence}. Here we fill this gap by testing the EDH for a paradigmatic case of  a heavy particle (subsystem $\cS$)  interacting with a gas of light particles (environment $\cB$). As established in the early days of the decoherence theory \cite{joos1985emergence,gallis1990environmental,Diosi_1995} and elaborated later on \cite{hornberger2003collisional,Hornberger_2006,vacchini2009quantum,hahn_2012_nonentangling,Lychkovskiy_2016},  a non-equilibrium spatially nonlocal state of a heavy particle rapidly turns into a mixture of well-localized classical-like states due to collisions with the particles of the gas (quantitative predictions of this theory were verified in interference experiments with fullerenes \cite{Hornberger_2003_collisional_experiment}).  Yet, surprisingly, we find that there exist non-classical eigenstates of the combined system $\cH$ with a coherence length on the order of the system size.

The rest of the paper is organised as follows. First we introduce the model. Then we introduce a figure of merit of the (non-)classicality -- the coherence length, and discuss how it can be used to test the EDH. Then we consider a trivially integrable case of an infinitely heavy particle. We demonstrate that in this case the EDH can be either valid or dramatically violated, depending on the spatial symmetry of the system. Next, we present the results of a numerical studies of a nonitegrable system. They strongly indicate the violation of the EDH. We conclude the paper by discussing the implications of this finding.

\medskip
\noindent{\it Model.}
We study a closed quantum system consisting of a distinguished particle (referred to simply as  ``the particle'' in what follows) interacting with the Fermi gas, the latter being regarded as an environment for the former.   For the sake of numerical calculations, we consider a one-dimensional lattice variant of this system with $N$ fermions and a single particle, all hopping on a linear lattice with $L$ sites.  The Hamiltonian reads
\begin{align}\label{H specific}
  H= & \left(-J\sum_{i=1}^{L-1} c_i^\dagger c_{i+1}-J'\sum_{i=1}^{L-1} a_i^\dagger a_{i+1}+h.c.\right)\nonumber\\
   & +  \sum_{i=1}^{L} U a_i^\dagger a_i c_i^\dagger c_i + \delta H.
\end{align}
Here  $c_i^\dagger$ creates a fermion and  $a_i^\dagger$  -- the particle, $J>0$ and $J'\geq0$ are hopping constants for fermions and the particle, respectively,  $U>0$ is the coupling constant between a fermion and the  particle. The term
\be
\delta H=\epsilon \sum_{j=1}^{L} (j/L) c_j^\dagger c_j
\ee
describes the linear on-site potential felt by fermions. It is introduced to break the otherwise present reflection symmetry of the Hamiltonian, for a purpose to be discussed later.
%and the last term describes ...
When considering large system size, we will imply the thermodynamic limit with the fixed fermionic density $n\equiv N/L$.
%The many-body eigenenergies and eigenstates  of this Hamiltonian are referred to as $E$ and $\Phi_E$, respectively, with $H\Phi_E=E\Phi_E$. %The Hilbert space $\cH$ of the model \eqref{H specific} depends on the number of fermions, $N$, and the number of lattice sites, $L$.

In order to separate time scales of decoherence and thermalization, we assume that the distinguished particle is heavy  as compared to the fermions \cite{lychkovskiy2011entanglement,vacchini2009quantum}, which amounts to  $J\gg J'$.

In the limiting case of the infinitely heavy particle, $J'=0$, the model becomes trivially integrable, with the eigenstates of the form
\be\label{eigenstate integrable}
\Phi_E=|j\rangle\otimes|F_E^j\rangle,
\ee
where $|j\rangle$ is the state of the particle localized on the site $j$ and  $|F_E^j\rangle$ is an  eigenstate of the $j$-dependent quadratic fermionic Hamiltonian
\begin{equation}\label{H_j}
  H_j=  \left(-J\sum_{i=1}^{L-1} c_i^\dagger c_{i+1}+h.c.\right) +   U c_j^\dagger c_j + \delta H.
\end{equation}
Note that if $\delta H=0$, the spectrum acquires degeneracies due to the reflection symmetry of  the Hamiltonian.

\medskip
\noindent{\it Coherence length and EDH.}
%A variety of quantumness measures has been proposed and used in different contexts, see a review \cite{frowis2018macroscopic}.
In order to formulate the EDH for a specific subsystem $\cS$, one needs to define a suitable measure of its quantumness.
%The eigenstate decoherence hypothesis applied to the model \eqref{H specific} asserts that $\densS_E$ is classical-like.
To this end we employ a natural measure of quantumness of the state $\densS$ of a point particle -- its coherence length \cite{Barnett_2000,Franke-Arnold_2001},
\be\label{coherence length}
l(\densS) \equiv
\sqrt{
\frac
{2\sum_{i,j=1}^{L} \big|\langle i | \densS|j\rangle\big|^2 (j-i)^2}
{\sum_{i,j=1}^{L} \big|\langle i| \densS|j\rangle\big|^2 }
}.
\ee
Here $\densS=\tr_{\cB} |\Psi\rangle\langle \Psi |$ is a reduced density matrix of the particle obtained from a pure state $\Psi$ of the closed particle-gas system and  $\langle i | \densS | j \rangle=\langle \Psi | a^\dagger_j a_i |  \Psi \rangle$ is its matrix elements  in the position basis.\footnote{Note that our definition \eqref{coherence length} is different from that in \cite{Barnett_2000} by a factor~$\sqrt{2}$.}
The coherence length effectively measures the spatial extension of a superposition of localized states, ranging from $0$ for a particle localized on a single site to $(L-1)$ for a highly non-classical state of the form
\be\label{nonclassical}
\Psi_Q=(1/\sqrt2)(a_1^\dagger+a_L^\dagger)|F\rangle,
\ee
where $|F\rangle$ is some state of $N$ fermions.

To simplify the terminology and notations, we will attribute coherence length also to pure many-body states, implying $l(\Psi)\equiv l\left(\tr_{\cB}|\Psi\ra\la \Psi|\right)$.

Importantly, while $l$ for nonclassical states can be on the order of the size of the system $L$, as in eq. \eqref{nonclassical}, for classical-like states  $l$ is independent on the system size. In particular, for the model \eqref{H specific} we expect that $l$ for classical-like states is bounded from above by the average interparticle distance $L/N$ times the probability for the scattering of a fermion off the particle. This expectation is grounded in understanding that each scattering event effectively entangles  the heavy particle with the fermionic environment and thus causes decoherence. Note that this physical picture implies a non-equilibrium process and may not apply to equilibrium states. We emphasize  that the hallmark of the EDH would be not some specific value of the coherence length of eigenstates accepted as classical (this is a matter of convention), but independence of the coherence length on the system size in the thermodynamic limit. The latter property is in the focus of the present study.
 %If $U\gg J$, the reflection probability is close to 1, and $l\sim n^{-1}$.

Analogously to the ETH \cite{Mori_2018}, one can discriminate between {\it strong} and {\it weak} forms of the EDH. The former concept implies that the EDH is valid for {\it all} eigenstates. In order to confirm the strong the EDH in our model, one needs to verify that the maximum over all eigenstates of the coherence length, $l_{\rm max}$, stays bounded when the system size grows. The weak EDH stands for a somewhat more vague conjecture that a vast majority of eigenstates satisfy the EDH. We will regard the weak EDH satisfied whenever the coherence length $l_{\rm av}$ averaged over all eigenstates remains bounded with  the  system size increasing.

\medskip
\noindent{\it EDH: integrable case.} In the integrable case, $J'=0$, each eigenstate  \eqref{eigenstate integrable} has zero coherence length and EDH is perfectly valid. However, in the case $\delta H=0$ the symmetry of the Hamiltonian leads to a caveat: all eigenstates of the form \eqref{eigenstate integrable} are doubly degenerate, and one can construct new eigenstates with a large coherence length. Consider e.g an eigenstate
\be\label{non-classical state}
\widetilde\Phi_E=\frac1{\sqrt2}\left( |1\rangle\otimes|F_E^1\rangle+|L\rangle\otimes|F_E^L\rangle\right),
\ee
where $|F_E^1\rangle$ and $|F_E^L\rangle$ are eigenstates of fermionic Hamiltonians $H_1$ and $H_L$, respectively. Its coherence length reads
\be
l(\widetilde\Phi_E)=(L-1) \, \big|\langle F_E^1 | F_E^L \rangle\big|^2 / \sqrt{1+\big|\langle F_E^1 | F_E^L \rangle\big|^2}.
\ee
The overlap $\big|\langle F_E^1 | F_E^L \rangle\big|^2$ entering this formula vanishes exponentially in the system size in the thermodynamic limit for most of the energy levels. For this reason the weak EDH is still satisfied. However, for some energy levels, in particular, for those corresponding to ground states of $H_1$ and $H_L$, the overlap  vanishes as slow as  $1/L^\alpha$, which is a manifestation of the Anderson orthogonality catastrophe~\cite{anderson1967infrared}. Importantly, the exponent $\alpha$ does not exceed unity~\cite{schonhammer1991orthogonality}. As a consequence,
\be
l(\widetilde\Phi_E)\sim L^{1-\alpha},\qquad \alpha<1
\ee
diverges in the thermodynamic limit. Thus we come to the conclusion that for an integrable Hamiltonian with a degenerate spectrum  some eigenstates have the coherence length on the order of the size of the system and thus violate the strong EDH spectacularly.\footnote{There is a very different type of anomalous, highly excited  eigenstates that violate the strong EDH. Namely, one can construct a many-body eigenstate of the free Fermion gas with zero densities at two distinct sites; the tensor product of this state and a superposition of the particle over these two sites is then an eigenstate of the entire Hamiltonian \eqref{H specific} with $J'=0$ and $\delta H=0$.  These states are specific for point interaction on a lattice and do not emerge for finite-range interactions, in continuum models or for finite mass of the particle, therefore we do not consider them in detail.}  This conclusion is particularly stunning in light of the fact that the integrable model under considerations is often employed in textbook demonstrations of the collisional decoherence \cite{schlosshauer2008decoherence}. However, in  these demonstrations  the decoherence process is viewed as a far-from-equilibrium dynamics described by a sequence of pairwise collisions of gas particles with the impurity particle. It is not very surprising that this approach may be inapplicable to equilibrium states in the bottom of the spectrum.

If one breaks the reflection symmetry of the Hamiltonian \eqref{H specific} by a nonzero $\delta H$, the degeneracy of the spectrum goes away, the construction \eqref{non-classical state} becomes impossible and no eigenstates with a nonzero coherence length emerge. The strong EDH is satisfied in this case.

Curiously, the Hamiltonian \eqref{H specific} with $\delta H=0$ has another integrable point, $J=J'$ \cite{mcguire1965interacting,castella1995integrability}. This case is superficial for studying the quantum-to-classical transition, since there is no separation of time scales of decoherence and thermalization, and there is no reason to expect the particle to be ``more classical'' than its environment. However, from the technical point of view, it would be instructive to check whether the coherence lengthes of some of eigenstates diverge for this another integrable point of the model. However, the explicit calculations in this model constitute a formidable challenge. Up to now, the only available analytical result is the particle's density matrix for the ground state in a translation-invariant system at a fixed total momentum, that implies a finite coherence length for this state~\cite{gamayun2019zero}.
% Thus it seems that the integrability on its own is not determinative for the validity of the EDH.

\medskip
\noindent{\it EDH: nonintegrable case.} To test the EDH in the nonintegrable case  we numerically diagonalize the Hamiltonian~\eqref{H specific} for finite system sizes and calculate the coherence length $l(\Phi_E)$  for each eigenstate $\Phi_E$. The range of system sizes is $L=8,9,\dots 12$. To keep the fermion density $N/L$ as independent on the system size as possible, we choose to half-fill the chain with fermions. Namely, we take $N=L/2$ for even $L$ and  $N=(L-1)/2$ for odd $L$. We consider both unbiased ($\varepsilon=0$) and biased ($\varepsilon=0.1$) versions of the Hamiltonian~\eqref{H specific}. Note that in the former case, despite the symmetry, the spectrum is nondegenerate.

\begin{figure}
\includegraphics[width = 0.4\textwidth]{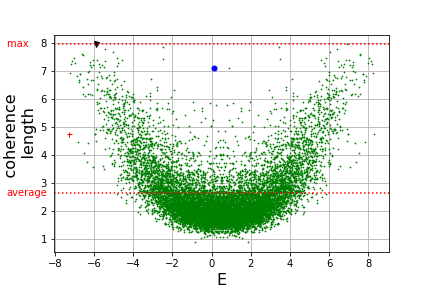}
\caption{The coherence length for eigenstates of the nonintegrable Hamiltonian \eqref{H specific} with $J=1,\,J'=0.2,\,U=1,\, N=6,\,L=12,\,\delta H=0$. The ground state, the state with the maximal coherence length and a state in the middle of the spectrum with a large coherence length are marked with the red cross, black triangle and blue  circle, respectively. The correlation function for these three states  is shown in Fig.~\ref{fig: correlation function}.}
\label{cohlength}
\end{figure}

\begin{figure}
\includegraphics[width = 0.4\textwidth]{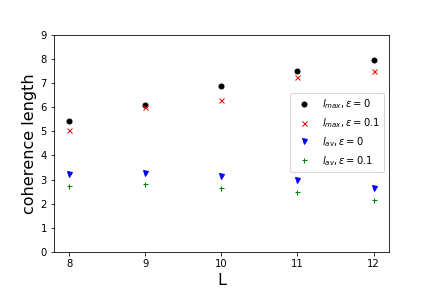}
\caption{The maximal  $l_{\rm max}$ and average $l_{\rm av}$ coherence lengths for eigenstates of the nonintegrable Hamiltonian for the cases of the presence ( $\epsilon=0$) and absence ($\epsilon=0.1$) of the spatial reflection symmetry. The system is half-filled with fermions, with $N=L/2$ for even $L$ and $N=(L-1)/2$ for odd $L$. Other parameters of the Hamiltonian are the same as in Fig.~\ref{cohlength}.
\label{scaling}}
\end{figure}

\begin{figure}
\includegraphics[width = 0.4\textwidth]{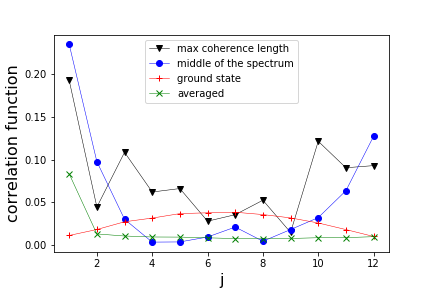}
\caption{Absolute value $ \big| \langle \Phi_E | a^\dagger_j a_1 |   \Phi_E\rangle \big|$ of a correlation function of a particle for three outlier eigenstates (highlighted in Fig.~\ref{cohlength}), compared to the same quantity averaged over all eigenstates. While the latter quantity rapidly decays at the length scale equal to the average distance between the fermions, the correlations for the outliers extend over the entire system. The parameters of the Hamiltonian are the same as in Fig. \ref{cohlength}.
\label{fig: correlation function}}
\end{figure}

The coherence lengthes of all eigenstates are presented in Fig. \ref{cohlength}. Most of the states have coherence length on the order of the average interparticle distance, $ L/N\simeq 2$, in accordance with expectation discussed above. In particular, the coherence length averaged over all states reads $l_{\rm av} = 2.63$. However, one can clearly see that there exists a number of "outlier" eigenstates with a large coherence length well exceeding this value. Such outliers are mostly concentrated at the edges of the spectrum, however they also can be found in the middle of the spectrum.

%It can be seen from Fig. \ref{cohlength}  that the coherence length is generally greater at the edges of the spectrum.  This tendency is consistent with the intuition based on the dynamical decoherence theory, since the edges of the spectrum correspond to a small in the absolute value (positive or negative) temperature, when the scattering processes should be suppressed due to the Pauli exclusion principle \cite{Hornberger_2006}.  However, even in the middle of the spectrum there are eigenstates with a large coherence length.

As discussed above, a decisive signature of violation of the EDH is the growth of the coherence length with the system size. In Fig.~\ref{scaling} we show the finite-size scaling of the average and maximal (over all eigenstates) coherence lengths. The scaling of these two quantities is drastically different: while the former vanishes with the system size, the latter grows linearly with $L$. Thus we conclude that our numerical data support the weak EDH, but suggest the violation of the strong EDH.

We have performed numerical calculations with various fermion densities (where we can access slightly larger system sizes, up to $L=12$), coupling strengths, as well as next-to-nearest neighbour interactions, and obtained similar results. Note that, in contrast to the integrable case, the potential bias $\delta H$ does not quantitatively alter the result, as is clear from Fig.~\ref{scaling}

To highlight the difference of outliers from typical eigenstates, we plot in Fig.~\ref{fig: correlation function} an absolute value  of a correlation function $ \langle \Phi_E | a^\dagger_j a_1 |   \Phi_E\rangle $ (as a function of the position $j$) for several outliers and compare it to the same quantity averaged over all eigenstates. One can see that while the averaged correlation function demonstrates a rapid decay at the length scale $L/N$, the correlation function of outliers extends over the entire system, even up to the opposite end of the chain.

%It can also be seen from Fig. \ref{cohlength}  that the coherence length is greater at the edges of the spectrum.  This is consistent with the intuition based on the dynamical decoherence theory, since the edges of the spectrum correspond to a small in the absolute value (positive or negative) temperature, when the scattering processes should be suppressed due to the Pauli principle \cite{Hornberger_2006}. This suppression of the scattering probability can be overridden by increasing the coupling between the particle and fermions. In the Supplement we show that this indeed happens.

%\noindent{\it Eigenstate decoherence {\it vs} eigenstate thermalization.}

%we note that decoherence is much less sensitive to integrability than thermalization. The reason is that the system is not able to explore whether it is integrable or not on the decoherence time scale. This implies that the extended eigenstate decoherence hypothesis should be applicable to  integrable systems without major modifications. We have tested that this is indeed the case in an integrable Heisenberg model.

\medskip
\noindent{\it Summary and discussion.}
To summarise, we have tested the eigenstate decoherence hypothesis (EDH) for a  system consisting of a heavy particle immersed in a one-dimensional Fermi gas. We have addressed the integrable case of an infinitely heavy particle as well as the  nonintegrable case of a particle of a finite mass. In the integrable case the EDH (both strong and weak versions) can be violated when the spatial symmetries of the model impose a massive degeneracy of the spectrum. On the other hand, if this degeneracy is lifted by a symmetry-breaking potential (integrability being preserved), the strong EDH is valid.

The most surprising results have been obtained numerically for a nonintegrable case. Our numerical data suggest that while the weak EDH holds, the strong EDH is violated by rare nonclassical outlier eigenstates with a particle coherence length on the order of the system size.

The existence of such outlier eigenstates is in stark contrast with the intuition based on the theory of collisional decoherence \cite{joos1985emergence,gallis1990environmental,Diosi_1995,hornberger2003collisional,Hornberger_2006,vacchini2009quantum}, which predicts a rapid decoherence as soon as the particle experience a collision with the gas particles. A plausible resolution of this conundrum is that most initial states (in particular, product states of a particle and a gas often considered in this theory) have a vanishing overlap with outliers, and thus the latter typically have no effect on the decoherence. A careful crafting of the state of a many-body system (e.g. in a cold atom simulator \cite{amico2020roadmap}) is required to unveil the nonclassical eigenstates.

Finally, we remark that the outlier nonclassical eigenstates violating the EDH resemble the many-body scars violating the ETH. One may wonder if the scars, in fact, exist in the model  \eqref{H specific}. At $J=J'$ and $\delta H=0$ the Hamiltonian~\eqref{H specific} is a Hubbard Hamiltonian, where scars do exist \cite{Yang_1989_eta-paiting,mark2020eta,moudgalya2020eta}. However, to the best of our knowledge, no scars have been reported away from this point. Furthermore, typically  the scars are not robust with respect to perturbations of the Hamiltonian \cite{lin2020slow}, while the violation of the EDH is. We therefore conclude that, anyway, the violation of EDH is not conditioned on the existence of quantum many-body scars.

%Remark about translation invariance

% Remark about MBL

%Remark about prospects of proof

%Product initial state vs correlated initial state

\medskip

\begin{acknowledgments}
{\it Acknowledgements.} OL is grateful to Oleksandr Gamayun and Anatoly Dymarsky for useful discussions.  This work is supported by the Russian Foundation for Basic Research under grant  18-32-20218.
\end{acknowledgments}

\bibliography{C:/D/Work/QM/Bibs/scars,C:/D/Work/QM/Bibs/1D,C:/D/Work/QM/Bibs/orthogonality_catastrophe,C:/D/Work/QM/Bibs/macrosuperpositions,C:/D/Work/QM/Bibs/decoherence,C:/D/Work/QM/Bibs/QIP,C:/D/Work/QM/Bibs/spin_chains,C:/D/Work/QM/Bibs/thermalization}
%\bibliography{supplement,1D,orthogonality_catastrophe,macrosuperpositions,decoherence,QIP,thermalization}

\end{document}